\documentstyle[12pt]{article}
\begin{document}
\title{
On Black Hole Creation in Planckian Energy Scattering
}

\author{ I.Ya. Aref'eva
 \thanks{Steklov Mathematical
Institute, Vavilov st.42, GSP-1, 117966, Moscow , Russia; E-mail :
arefeva@arevol.mian.su}, K.S. Viswanathan
 \thanks{Department of Physics, Simon Fraser University,
Burnaby, British Columbia, V5A 1S6, Canada,
 E-mail: kviswana@sfu.ca}
and I.V.Volovich\thanks{Steklov Mathematical
Institute, Vavilov st.42, GSP-1, 117966, Moscow , Russia; E-mail:
volovich@mat.utovrm.it}}
\date {$~$}
\maketitle
\begin{abstract}

In a series of papers Amati, Ciafaloni and Veneziano and 't Hooft
conjectured that black holes occur in the collision of two light particles
at planckian energies. In this talk based on  \cite {AVV}
we discuss a possible scenario for such a process by using
the Chandrasekhar-Ferrari-Xanthopoulos duality between the Kerr
black hole solution and  colliding plane gravitational waves.

\end{abstract}

\vspace{1cm}
\section{Introduction}
\setcounter{equation}{0}

An idea by M.A.Markov \cite {Mar,Mar95}, see also \cite {MF}, that
black holes can be considered as elementary particles becomes
nowadays a common place in quantum gravity \cite {Haw}  and
superstring theory \cite {Sen,Wit,Hall}.
If black holes are particles then one can ask about creation and
annihilation of black holes in scattering processes
of ordinary particles with very high energies.
Amati, Ciafaloni and Veneziano \cite{1} and 't Hooft \cite{2} made
a conjecture  that black holes
will occur in the collision of two light particles at Planck energies
with small impact parameters.

In this talk based on  \cite {AVV} we discuss the following  possible
mechanism
of black holes creation

$$\mbox {{\bf Particles }}~\to ~
\mbox {{\bf gravitational waves}}~\to ~
 \mbox {{\bf Black Holes}}
$$

Ultra-relativistic particles generate  plane gravitational waves
then these plane gravitational waves collide and
produce a singularity or a black hole.

We are going to discuss some issues concerning quantum mechanical
description  of black hole creation in the scattering processes of particles.
Note that black holes  cannot be incorporated into the
theory if we consider quantum field theory in  Minkowski space-time.

It is difficult to perform
calculations for such a process in a realistic situation. We shall
discuss here an idealized picture.  Any gravitational wave far away from
sources can be considered as a plane wave. We assume that plane
waves already have been produced by ultrarelativistic particles
and then consider analytically the process of black hole formation
under interaction of these plane waves.

We discuss  the semiclassical transition amplitude for the process
of creation of black hole in the collision of two plane waves
in  the semiclassical approximation.
For this purpose use a classical discribtion
of the process

$$\mbox {{\bf gravitational waves}}~\to ~
 \mbox {{\bf Black Holes}}
$$

Classical collision of plane gravitational waves has been the
subject of numerous investigations, see for example
\cite{SKP,CX,FI,Gr}, and it has a remarkably rich structure.
Here we are going to
use the Chandrasekhar-Ferrari-Xanthopoulos duality between
colliding plane
gravitational waves and the Kerr black hole solution.

One has two dimensionless parameters in a scattering process at Planck
energies\cite{1,2}. If $E$ is the energy in the center of mass frame, then
one defines the Planckian energy regime to be $GE^{2}/\hbar\geq 1$, where
$G$ is the Newton constant. This means that one can treat the process
semiclassically. The other dimensionless parameter is $GE/b$, where $b$
is the impact parameter. If one takes $GE/b\ll1$ , one can use the eikonal
approximation. The elastic scattering amplitude $A(s,t)$ in the eikonal
approximation was found in \cite{1,2}. Fabbrichesi, Pettorino,
Veneziano and Vilkovisky (FPVV) \cite{7}
gave the representation
\begin{equation}
A(s,t) \sim s \int d^{2}b~ e^{iqb} e^{iI_{cl}(s,b)}.
\end{equation}
Here $s$ and $t$ are the Mandelstam variables, $q^{2}=-t$.
$I_{cl}(s,b)$ was taken to be the value of the
boundary term for the gravitational
action calculated on the sum of
two Aichelburg-Sexl shock waves,
\begin{equation}
I_{cl}(s,b)=G s \log b^2.
\end{equation}
This corresponds to a linearization of
the problem and one cannot see in such
an approximation black hole formation.
It is tempting to suggest that one
can use the FPVV approach  \cite{7} and
the formula of type (1) even
for small impact parameters to calculate
the phase of the transition
amplitude from a state containing two
particles to a state containing a black hole. In such a case
$I_{cl}(s,b)$ could be the value
of the gravitational action for a
solution of Einstein equation
corresponding to an interior of the Kerr
black hole created in the collision of
plane gravitational waves. The
parameters $s$ and $b$ in this case
could be expressed in terms of the
mass and angular momentum  of the Kerr black hole.

\section{Black Hole   Transition Amplitude }
 \setcounter{equation}{0}

We discuss here some issues concerning quantum mechanical
description
of black hole creation in the scattering processes of particles.
Let us note first that black holes  cannot be incorporated into the
theory if we consider quantum field theory in  Minkowski space-time.
In fact it is obvious
from Einstein
equation
\begin{equation}
        R_{\mu \nu }-\frac{1}{2}g_{\mu \nu}R=8\pi GT_{\mu \nu}
	\label{2.1}
\end{equation}
that if one has  matter, i.e. a nonvanishing $T_{\mu \nu}$
then one has a nontrivial gravitational field. This means that
one has to
start with a region of space-time which is flat only in some
approximation. Then, in the process of collision, one gets a
strong gravitational field including perhaps black holes and/or
singularities.
Let us clarify the
meaning of the transition amplitude from a state
describing particles to a
state containing black holes,
$$<\mbox {Black holes}|\mbox{Particles in almost
 Minkowski space-time}>.$$

 By  analogy with the definition
 of transition amplitude in quantum mechanics
this transition amplitude
can be characterized  by values of the metric and others fields in two
given moments of time (or by data on two given Cauchy surfaces,
say $\Sigma '$ and $\Sigma ''$).

Our starting point is the quantum-mechanical Feynman transition
amplitude between  definite configurations of the three-metric $h_{ij}'$
and field $\Phi '$ on an initial spacelike surface  $\Sigma ' $
and a configuration $h_{ij}''$ and  $\Phi ''$ on a final surface
 $\Sigma '' $. This is
 \begin{equation}
<h'',\phi '', \Sigma ''|h', \phi ' , \Sigma '>=
\int ~~ e^{\frac{i}{\hbar}S[g,\Phi]}
{}~{\cal D}\Phi {\cal D}g,                               \label {3}
\end{equation}
where the integral is over all four-geometries and field configurations
which match  given values on two spacelike surfaces, i.e.
 \begin{equation}
\Phi |_{\Sigma '}=\phi ',~
g |_{\Sigma '}=h '                                    \label {4}
\end{equation}
 \begin{equation}
\Phi |_{\Sigma ''}=\phi '',~
g |_{\Sigma ''}=h ''                                    \label {5}
\end{equation}
In this paper we shall consider the transition amplitude
in the semiclassical approximation and we don't
discuss here the introduction of ghosts and the
definition of the measure $Dg$.

We are interested in the process of black hole creation. Therefore we
specify the initial configuration $h' $ and $\phi ' $ on $\Sigma '$
as configuration of gravitational and matter fields in Minkowski
spacetime and we specify the final configuration $h''$ and $\phi ''$
on $\Sigma ''$
as describing a black hole.
So, $\Sigma '$ {\it is a partial Cauchy surface with
 asymptotically simple
past in a strongly asymptotically predictable space-time and }
$\Sigma ''$ {\it is a partial Cauchy surface
 containing black hole(s), i.e.
$\Sigma ''-J^{-}({\cal T}^{+})$ is non empty}.
To explain these let us recall some necessary notions
from the theory of black holes \cite {HE,Wa}.

Black holes  are conventionally defined in
asymptotically flat space-times by the
existence of an event horizon $H$.
The horizon $H$ is the boundary $\dot {J}^{-}({\cal I}^{+})$
 of the causal past $J^{-}({\cal I}^{+})$
of future null infinity ${\cal I}^{+}$, i.e. it is the boundary of the
set of events
in space-time from which one can escape to infinity in the future
direction.

The {\it black hole region } $B$ is $$B=M-J^{-}({\cal I}^{+})$$
and {\it the event horizon}
$$H=\dot {J}^{-}({\cal T}^{+}).$$

Consider a space that is asymptotically flat in the sense of being
{\it weakly asymptotically simple and empty}, that is, near future and
past null infinities it has a conformal structure like that of Minkowski
space-time. One assumes that  space-time is {\it future asymptotically
predictable}, i.e. there is a surface ${\cal S}$ in spacetime that serves
as a Cauchy surface for a region
extending to future null infinity. This means that there are no "naked
singularities"
(a singularity that can be seen from infinity)
to the future of the surface ${\cal S}$.
This gives a formulation of Penrose's cosmic censorship
conjecture.

A space-time $(M, g_{\mu \nu})$
is {\it asymptotically simple} if there
exists a smooth manifold $\tilde {M}$
with metric $\tilde {g}_{\mu\nu}$,
boundary ${\cal T}$, and a scalar
function  $\Omega $ regular everywhere on
$\tilde {M}$ such that

(i) $~~\tilde {M}-{\cal I}$ is conformal
to $M$ with  $\tilde {g}_{\mu\nu}=
\Omega ^{2} g_{\mu\nu}$,

(ii) $~\Omega >0$ in $\tilde {M}-{\cal T}$
and     $\Omega =0$ on ${\cal I}$
with $\nabla _{\mu }\Omega \neq 0$ on ${\cal I}$,

(iii) Every null geodesic
on $\tilde {M}$ contains, if maximally extended,
two end points on  ${\cal I}$.

If $M$ satisfies the Einstein vacuum equations near ${\cal I}$
then ${\cal I}$ is null. ${\cal I}$
consists of two disjoint pieces ${\cal I}
^{+}$ (future null infinity)
and ${\cal I}^{-}$ (past null infinity)
each  topologically is $\approx R$x$S^{2}$.

A space-time $M$ is {\it weakly asymptotically simple}
 if there exists
an asymptotically simple $M_{0}$ with corresponding $\tilde{M}_{0}$
such that for some open subset K of  $\tilde{M}_{0}$
including ${\cal I}$,
the region $M_{0}\cup K$ is isometric to an open
 subset of $M$.  This
allows $M$ to have more infinities than just ${\cal I}$.

The domain of dependence $D^{+}(\Sigma)$ ($D^{-}(\Sigma)$)
of a set $\Sigma$ is defined as the set of all points $p\in M$
such that every  past (future) inextendible non-spacelike
 curve through $p$
intersects $\Sigma$. A space like hypersurface which no non-spacelike
curve intersects more than once is called a partial Cauchy surface.
Define $D(\Sigma) =D^{+}(\Sigma)\cup D^{-}(\Sigma)$.
A partial Cauchy surface $\Sigma$ is said to be a global
 Cauchy surface
if $D(\Sigma)=M$.

Let $\Sigma $ be  a partial Cauchy surface in a weakly
 asymptotically simple
and empty space-time $(M,g)$. The space-time $(M,g)$ is (future) {
\it asymptotically
predictable from} $\Sigma$ if ${\cal I}^{+}$ is contained
in the closure
of $D^{+}(\Sigma)$ in $\tilde{M}_{0}$.
If, also, $J^{+}(\Sigma) \cap \bar{J^{-}}({\cal I}^{+},\bar{M})$
is contained in $D^{+}(\Sigma)$ then the space-time $(M,g)$
is called strongly   asymptotically predictable  from $\Sigma$.
In such a space there exist a family $\Sigma (\tau)$, $0<\tau<\infty$,
of spacelike surfaces homeomorphic to $\Sigma$ which cover
$D^{+}(\Sigma)-\Sigma$ and intersects ${\cal I}^{+}$.
 One could regard
them as surfaces of constant time.
A {\it black hole on the surface} $\Sigma (\tau)$ is
 a connected component of the set
$$
B(\tau)=\Sigma (\tau)- J^{-}({\cal I}^{+},\bar{M}).$$

One is interested primarily in
black holes which form from an initially
non-singular state. Such a state
can be described by using the partial
Cauchy surface $\Sigma$ which has an {\it asymptotically simple past},
i.e.
the causal past $J^{-}({\Sigma})$ is isometric to the region
 $J^{-}(\cal I)$ of some asymptotically
 simple and empty space-time with
 a Cauchy surface ${\cal I}$. Then $\Sigma$ has the topology $R^{3}$.

In the case considered one has a space-time
$(M,g_{\mu\nu})$ which is weakly asymptotically simple and empty
and  strongly asymptotically
predictable  .

$\Sigma ' $ is a partial Cauchy surface with asymptotically simple past,
$\Sigma ' \sim R^{3}$.

 $\Sigma ''=\Sigma (\tau '') $ contains a black hole, i.e. $\Sigma '' -$
$J^{-}({\cal I} ^{+}, \bar{M})$ is nonempty.

In particular, if one has the
condition $\Sigma ' \cap \bar {J}^{-}({\cal I})$
is homeomorphic to $R^{3}$ (an open set with compact closure)
then $\Sigma ''$ also satisfies this condition.

Strictly speaking one cannot apply the standard definition of black holes
to the case of  plane gravitational waves and we need a
generalization.One defines a black hole in terms of the event horizon,
 $\dot {J}^{-}({\cal I} ^{+})$.
However this definition depends on
the whole future behaviour of the metric.
There is  a different sort of horizon
which depends only on the properties
of space-time on the surface  $\Sigma (\tau) $ \cite {HE}.
Any point in the black hole region bounded by $r=2m$ in the Kruskal
diagram represents a {\it trapped surface} (which is a two-dimensional
sphere in space-time) in that both the outgoing and ingoing families of
null geodesics emitted from this point converge and hence no light ray
comes out of this region. A generalization of the definition of black
holes in terms of trapped horizon has been considered in \cite
{MT,Hay}.
A generalization of the standard definition of black holes to the case of
nonvanishing cosmological constant was considered by Gibbons and Hawking
[15].

We discussed the transition amplitude (propagator) between definite {\it
configurations} of fields,
$<h'',\phi '', \Sigma ''|h', \phi ' , \Sigma '>$.
The transition amplitude from {\it a state} described by the wave
function $\Psi ^{in}[h', \phi ']$ to a state $\Psi ^{out}[h'', \phi '']$
reads
\begin{equation}
<\Psi ^{out}|\Psi ^{in}>=
\label {7}
\end{equation}
$$\int \bar{\Psi }^{out}[h'', \phi '']
<h'',\phi '', \Sigma ''|h' ,\phi ' , \Sigma '>\Psi ^{in}[h' ,\phi ']
{\cal D}h'{\cal D}\phi '{\cal D}h''{\cal D}\phi ''.$$
One can take for example the state $\Psi ^{in}=
\Psi ^{in}_{p_{1}p _{2}}[h', \phi ']$
as a Gaussian distribution describing a state of particles with momenta
$p_{1}$ and $ p_{2}$ and take $\Psi ^{out}$ as a wave function describing a
state of black hole. Recently Barvinski, Frolov and Zelnikov
have suggested an expression for the wave function of the ground state of a
black hole \cite {Frolov}.

\section{Boundary term in Gravitational Action
and Semi\-classical Expansion }
\setcounter{equation}{0}

In this section we discuss an approximation scheme for calculating
the transition amplitude following the FPVV approach \cite{7}.
The gravitational action with the boundary term has the form \cite{York}
\begin{equation}
S[g]=-\frac{1}{16\pi G}\int_{V} R\sqrt{-g}d^4x -
\frac{1}{8\pi G}\int_{\partial V} K\sqrt{h}d^3x \label {3.1}
\end{equation}
Here $V$ is a domain in space-time
with the space-like  boundary   $\partial V$,
$h$ is the first fundamental form and $K$ is the trace of the
second fundamental form of $\partial V$.

The case of null surfaces was considered in \cite {Isr}.
We shall
write a representation of the
boundary term suitable for quantum consideration.
The action is
 \begin{equation}
S[g]=-\frac{1}{16\pi G}\left(\int_{V} d^{4}x \sqrt{-g}~ R (g)+
\int _{V}d^{4}x \sqrt{-g}~\nabla _{\mu}f^{\mu}(g)\right),
                                                      \label {3.2}
\end{equation}
where
 \begin{equation}
                                                       \label {3.3}
f^{\mu}(g)=g^{\alpha\beta}g^{\mu \nu}
\partial_{\nu} g_{\alpha\beta}-
g^{\mu\alpha}g^{\beta \nu}\partial_{\nu}g_{\alpha\beta}.
\end{equation}
Supposing that the boundary is described by equation
 \begin{equation}
 \sigma (x) =0
                \label {3.10}
\end{equation}
one gets the action in the form
 \begin{equation}
                \label {3.11'}
S[ g]=-\frac{1}{16\pi G}(\int _{V}d^{4}x \sqrt{-g}~ R (g)+
\int _{V}d^{4}x \sqrt{-g_{cl}}~\delta (\sigma (x))~
f^{\mu}\nabla_{\mu}\sigma ).
\end{equation}

The linearization of the action (\ref {3.2}) leads to the action
in the FPVV form  \cite {7}.

Let us show that the presence of the boundary term in the action
(\ref {3.2}) is necessary for the selfconsistency of the semiclassical
expansion. To perform semiclassical
expansion, one expands the metric $g$ around a
classical solution $g_{cl}$ of the Einstein
equation so that $g=g_{cl}+\delta g$,
 \begin{equation}
S[g]=S[g_{cl}+\delta g]=S[g_{cl}]+S'[g_{cl}]\delta g
+\frac{1}{2}S''[g_{cl}](\delta g)^{2}+...,
\label {3.3'}
\end{equation}
where $g_{cl}$ matches the given Cauchy data on surfaces
$\Sigma '$ and $\Sigma ''$, i.e.
 \begin{equation}
g  _{cl}~|_{\Sigma '}=h ' ,~~\label {3.4}
g  _{cl}~|_{\Sigma ''}=h ''
\end{equation}
In this case $\delta g|_{\Sigma '}=\delta g|_{\Sigma ''}=0$, but we
cannot guarantee that $\nabla\delta g|_{\Sigma '}=\nabla\delta g|_{\Sigma
''}=0$.
To ensure that terms linear in $\nabla\delta g$
drop out from expression (\ref{3.3'}) (otherwise we cannot perform
semiclassical expansion) one has to integrate by parts.  One has
\begin{equation}
S[ g +\delta g]=-\frac{1}{16\pi G}\left(\int d^{4}x
\sqrt{-g }~R
(g )+
\int d^{4}x \sqrt{-g }~\nabla _{\mu}f^{\mu}(g )) + \right.
 \label {3.8}
 \end{equation}
$$+\left. \int d^{4}x \sqrt{-g }~(g^{\mu \nu}
\nabla ^{2}\delta g_{\mu \nu}
-\nabla ^{\mu}\nabla ^{\nu}\delta g_{\mu \nu})+
\nabla _{\mu}( g^{\alpha\beta}\partial^{\mu}\delta g_{\alpha\beta }
 -g^{\mu\alpha}\partial ^{\beta}\delta g_{\alpha\beta })\right)
$$
$$+~ second ~~order~~terms.
$$
In eq.(3.8) we have dropped the subscript $cl$ from $g_{cl}$.
The linear terms coming from  the Hilbert-Einstein action
 can be put in the form
 \begin{equation}
                          \label {3.9}
g^{\mu \nu} g^{\alpha\beta}\nabla _{\alpha}
\nabla _{\beta}\delta g_{\mu \nu}
-\nabla ^{\mu}\nabla ^{\nu}\delta g_{\mu \nu}=
\nabla _{\mu}( g^{\alpha\beta}\nabla^{\mu}\delta g_{\alpha\beta }
 -g^{\mu\alpha}\nabla ^{\beta}\delta g_{\alpha\beta })
\end{equation}
Notice that on the RHS of the last relation covariant derivatives
can be replaced by partial  derivatives because
on the boundary $\delta g =0$.
Therefore one finds that terms linear in
$\delta \partial g $ coming from Hilbert-Einstein action
cancel  similar terms coming from
full divergence and the action (\ref {3}) admits
the expansion (\ref {3.3'}).

Taking into account that the value of the Hilbert-Einstein action
 on the classical
solution is equal to zero, one finds that the full action for a solution of
Einstein equation is
reduced to the second term in (\ref{3.8}) which
can be reduced to a boundary term.

The transition amplitude (\ref{3}) in semiclassical approximation is
 \begin{equation}
<h'',\Sigma ''|h' ,\Sigma '>={\cal Z} \exp {\frac{i}{\hbar }S_{cl}}
\label {3.12}
\end{equation}
where
 \begin{equation}
                \label {3.11}
S[ g_{cl}]=-\frac{1}{16\pi G}\int d^{4}x \sqrt{-g_{cl}}~\delta (\sigma (x))~
f^{\mu}\nabla_{\mu}\sigma ,
\end{equation}
 $f$ is given by equation (\ref{3.3}) and
$$
{\cal Z}=(\frac{\pi}{\det S''(g_{cl})})^{1/2}
$$
We assume here that there is only one solution
of classical equation of motion with given boundary conditions.

\section{Colliding Plane Gravitational
Waves }
\setcounter{equation}{0}

There exists a well known class of plane-fronted gravitational waves
with the metric
\begin{equation}
        ds^{2}=2dudv +h(u,X,Y)du^{2}-dX^{2}-dY^{2}
	\label{o}
\end{equation}
where $u$ and $v$ are null coordinates .
In particular the gravitational field
of a particle moving with the speed of
light is given by the Axelburg-Sexl solution. The metric has the form
\begin{equation}
ds^{2}=2dudv + E\log(X^2+Y^2)\delta(u)du^{2}-dX^{2}-dY^{2}
        \label{oo}
\end{equation}
and describes a shock wave. It is difficult to find a solution which
describes two sources.  An approximate solution of
equation (\ref{2.1}) for two particles as the sum of solutions each of
which describes one particle was considered by
FPVV\cite {7}. This
approximation describes well the scattering amplitude for large impact
parameter, but does not describe non-linear interaction of shock waves
which is dominant in the region of small impact parameter. To analyze
non-linear effects we will, instead of dealing with shock waves,  take
a  simple solution of Einstein equation, namely we will
take  plane gravitational
waves. In some respects
one can consider plane wave as an approximation to  more complicated
gravitational waves, in particular shock waves.
This solution in some sense can be interpreted as an approximation for a
solution of Einstein equation in the presence of two
moving particles.
Collision of two ultrarelativistic black holes was considered
by D'Eath \cite{DE}.
If our particles are gravitons then there are no
sources corresponding to matter fields in Einstein equations.
Note also that plane gravitational waves are produced
by domain walls, see \cite {4}.

\begin{figure}
\setlength{\unitlength}{0.5cm}
\begin{center}
\begin{picture}(15,10)(-5,-7)
  \put(-2.5,2.5){\line(1,0){5.0}}
  \multiput(-2.5,2.5)(0.25,0.0){21}{\circle*{0.2}}
\put(-5.0,-5.0){\vector(1,1){10.0} }
\put(-4.5,-4.5){\vector(1,1){2.0} }
\put(4.5,-4.5){\vector(-1,1){2.0} }
\put(5.0,-5.0){\vector(-1,1){10.0} }
        \put(-10.0,-5.0){\line(1,1){7.5}}
 \put(-10.0,-5.0){\line(1,1){7.0}}
\put(10.0,-5.0){\line(-1,1){7.5}}
\multiput(0.0,0.0)(0.0,0.5){5}{\vector(0,1){0.3}}
\multiput(0.5,0.5)(0.0,0.5){4}{\vector(0,1){0.3}}
\multiput(1.0,1.0)(0.0,0.5){3}{\vector(0,1){0.3}}
\multiput(1.5,1.5)(0.0,0.5){2}{\vector(0,1){0.3}}
\multiput(2.0,2.0)(0.0,0.5){1}{\vector(0,1){0.3}}
\multiput(-0.5,0.5)(0.0,0.5){4}{\vector(0,1){0.3}}
\multiput(-1.0,1.0)(0.0,0.5){3}{\vector(0,1){0.3}}
\multiput(-1.5,1.5)(0.0,0.5){2}{\vector(0,1){0.3}}
\multiput(-2.0,2.0)(0.0,0.5){1}{\vector(0,1){0.3}}

 \put(5.5,4.0){\mbox {$v$}}
 \put(-5.5,4.0){\mbox {$u$}}
\put(-4.5,-2.0){\mbox {{\bf II}}}
\put(3.5,-2.0){\mbox {{\bf III}}}
\put(0.0,-5.5){\mbox {{\bf I}}}
\put(-0.5,1.2){\mbox {{\bf IV}}}
\end{picture}
 \end{center}
\caption{$(u,v)$ plane wave coordinates }\label{f1}
\end{figure}
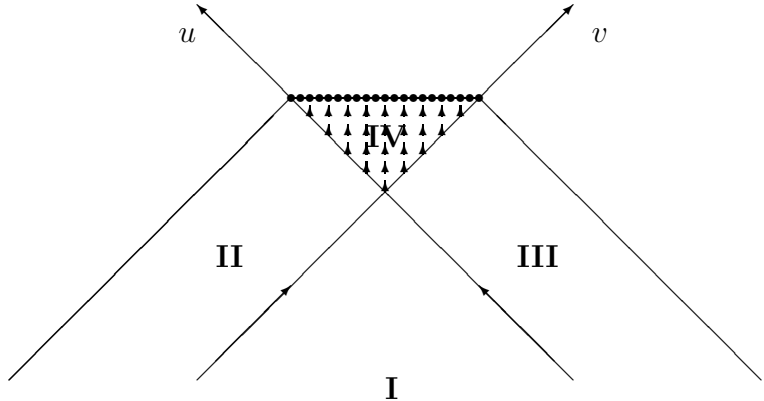

A particular class of
{\it plane waves} is defined to be
plane-fronted waves in which the field components are the same at every
point of the wave surface. This condition requires that $h(u,X,Y)$
is a function with a quadratic dependence on $X$ and $Y$.
One can then remove the dependence of $h$ on $X$ and $Y$ altogether by
a coordinate change.
Solutions of Einstein equations describing {\it collision } of
plane gravitational waves
were first obtained by Szekeres  and Khan and Penrose
\cite{SKP}.  Chandrasekhar, Ferrari and Xanthopoulos
\cite {CX,FI} have developed a powerful method for obtaining such solutions
by using a remarkable analogy ("duality") with stationary axisymmetric
case  which
 can be reduced to the investigation of the Ernst equation. For
a review see the Griffiths book \cite {Gr}.

We will use the coordinates $(u,v,x,y)$.
We assume that throughout  space-time there exists a pair of commuting
space-like Killing vectors $\xi _{1}=\partial _{x}$,  $\xi _{2}=\partial
_{y}$ . The Szekeres line element has the form
\begin{equation}
ds^{2}=2e^{-N}dudv -e^{-U}(e^{V}\cosh Wdx^{2}-
2\sinh W dxdy+e^{-V}\cosh Wdy^{2}),
	\label{4.3}
\end{equation}
Here $N,U,V$ and $W$ are functions of $u$ and $v$ only.

We illustrate in Fig.1 the two-dimensional geometry of plane waves.
Space-time is divided into four regions. The region {\bf I}
is the flat background before the arrival of the plane waves.
The null hypersurfaces $u=0,$
and $v=0$ are the past wave fronts of the incoming plane waves 1 and 2.
The metric in
region {\bf I} is Minkowski. Regions {\bf II} and {\bf III}
represent incoming plane waves which interact
in  region {\bf IV}.  Colliding plane gravitational waves can
produce singularities or Cauchy horizons in the interaction region
\cite{MT,CX,Yu1,KH,Hay}. The solution is undetermined
across a Cauchy horizon \cite{CX,Yu2,Hay,HEr} into the future. We shall discuss
two simplest extensions.

In particular, one can get an interior of the Schwarzschild solution
in the interaction region {\bf IV}.There are two types of colliding
plane waves solutions corresponding to the Schwarzschild metric. The
first one creates the interior of the black hole with the usual
curvature singularity. In this case incoming plane waves have curvature
singularities already before collision. In the context of the Planck
energy scattering it seems more natural that we don't have curvature
singularities
already for free plane gravitational waves. Therefore we will be
discussing mainly another type of solutions one gets in the interaction
region, namely,
the interior of the Schwarzschild white hole. The maximal analytic extension
of this solution across its Killing-Cauchy horizon leads to creation of a
covering space of the Schwarzschild black hole out of collision between
two plane gravitational waves. An alternative interpretation of this
solution is the creation of the usual Schwarzschild black hole
out of the collision between two plane gravitational waves propagating in
a cylindrical universe. There exists also a time-reversed extension
\cite {Hay} including the covering space of the Schwarzschild exterior
and part of black hole, and giving two receding plane waves with flat
space between. We will interpret this as scattering of plane waves
on the virtual black hole.

Vacuum Einstein equations
for the metric (\ref{4.3}) give a system of differential equations
for functions $U$, $V$, $N$ and $W$. One of these equations
can be integrated to give (see  \cite {AVV} for details)
\begin{equation}
 e^{-U} =f(u)+g(v),
        \label{4.12'}
\end{equation}
where $f(u)$ and $g(v)$ are arbitrary functions. We fix a gauge by using
$f=f(u)$ and $g=g(v)$ as new coordinates instead of $u$ and $v$ and then
we change variables from $f$
and $g$ to $\mu $ and $\eta$ such that
\begin{equation}
f+g= (1-\mu ^{2})^{1/2} (1-\eta ^{2})^{1/2},~~f-g=-\mu\eta .
	\label{4.13}
\end{equation}

By introducing the complex valued function
\begin{equation}
Z=\chi +i\lambda,
	\label{4.12}
\end{equation}
where
\begin{equation}
\chi =e^{-V}/ \cosh W , ~~  \lambda = e^{-V}\tanh W       .
	\label{4.11}
\end{equation}
one can reduce the system of differential equations
for functions $U$, $V$, $N$ and $W$ to the Ernst equation on $Z$.
In particular, this procedure gives the following solution of the
vacuum Einstein equations
\begin{equation}
ds^{2}= X(
\frac{d \eta ^{2}}{1-\eta ^{2}}-\frac{d \mu ^{2}}{1-\mu ^{2}})-
(1-\mu ^{2})^{1/2}(1-\eta ^{2})^{1/2}[\chi dy^{2}+\frac{1}{\chi}
(dx-\lambda dy)^{2}],
        \label{43.9}
\end{equation}
where
\begin{equation}
        \chi   =  (1-\mu ^{2})^{1/2}(1-\eta ^{2})^{1/2}
        \frac{X}{Y}
        \label{43.8}
\end{equation}
\begin{equation}
\lambda   =  \frac{2q}{p}[\frac{1}{1+p }-
  \frac{(1-\eta ^{2})(1-p\mu)}{1-p ^{2}\mu ^{2}-q^{2}\eta ^{2}} ]
        \label{43.8'}
\end{equation}
and
\begin{equation}
X= (1-p\mu) ^{2}+q^{2}\eta ^{2},~~~Y= 1-p ^{2}\mu ^{2}-q^{2}\eta ^{2},
	\label{43.10}
\end{equation}
$p$ and $q$ satysfy
        \begin{equation}
        E= p \mu  +iq \eta .
		\label{43.5}
        \end{equation}

 Now let us take
\begin{equation}
p=-\frac{(m^{2}-a^{2})^{1/2}}{m}, ~~ q= \frac{a}{m}, ~ m\geq a
	\label{43.11}
\end{equation}
 and introduce the new coordinates
 $(t,r,\theta , \phi )$ instead of $(\mu, \eta ,x,y)$ by putting
 \begin{equation}
\mu =\frac{r-m}{(m^{2}-a^{2})^{1/2}}, ~~ \eta =\cos \theta
 	\label{43.12}
 \end{equation}
 \begin{equation}
t= -\sqrt{2} m(x-\frac{2q}{p(1+p)}),~~ \phi =\frac{\sqrt{2} m}
{(m^{2}-a^{2})^{1/2}}y
        \label{4.13'}
\end{equation}
Then the metric (\ref {43.9}) will take the form of the Kerr solution
\begin{equation}
2m^{2}ds^{2}= (1-\frac{2mr}{\rho ^{2}})dt ^{2}-\frac{4amr}{\rho ^{2}}
\sin ^{2}\theta dt d\phi -
(r^{2}+ a ^{2}-\frac{2a^{2}mr}{\rho ^{2}})\sin ^{2}\theta  d\phi ^{2}
-\rho ^{2}(\frac{1}{\Delta} dr^{2} + d \theta ^{2})	,
	\label{43.14}
\end{equation}
where
\begin{equation}
\rho ^{2}= r^{2}+ a ^{2} \cos ^{2 }\theta, ~~\Delta= r^{2}-2mr+ a ^{2}
	\label{43.15}
\end{equation}
The coordinates must satisfy the inequality $|\eta|<\mu\leq 1$. This
implies that $-(m^{2}-a^{2})\sin ^{2}\theta <\Delta \leq 0 $, which means
the region of the Kerr spacetime that is inside the ergosphere.

To describe
the colliding plane
gravitational waves producing
the interior of the ergosphere in  the Kerr spacetime it is  convenient to
rewrite the metric (\ref {43.9}) in terms of the $(u',v')$ coordinates
related with $(\eta ,\mu )$ by the following relations
\begin {equation} 
                                                          \label {43.16}
\eta =\sin (u'-v'),~~\mu =\sin (u'+v'),
\end   {equation} 
These $(u',v')$ are related to $(u,v)$ in (\ref {4.3}) by $u=\sin u'$,
$v=\sin v'$. For simplicity of notations we will omit $'$ in (\ref {43.16}).
We have
\begin{equation}
ds^{2}= 4X(u,v)dudv-\Omega (u,v)[\chi (u,v)dy^{2}+\frac{1}{\chi (u,v)}
(dx-\lambda (u,v) dy)^{2}],
        \label{43.17}
\end{equation}
where
\begin {equation} 
                                                          \label {43.18}
X(u,v)=(1-p\sin (u+v))^{2}+q^{2}\sin ^{2}(u-v),~~\Omega (u,v)=
\cos(u+v)\cos(u-v),
\end   {equation} 
\begin {equation} 
                                                          \label {43.19}
\lambda (u,v)=\frac{2q}{1+p}\frac{1-\sin (u+v) }{Y(u,v)}((p+1)\sin ^{2}(u-v)+
p\sin (u+v)-1)
\end   {equation} 

\begin {equation} 
                                                          \label {43.20}
Y(u,v)=1-p^{2}\sin ^{2}(u+v)-q^{2}\sin ^{2}(u-v).
\end   {equation} 
In (\ref {43.17})
\begin{equation}
0<u<\pi/2,~~0<v<\pi/2,~~v+u<\pi/2
                                                          \label{43.21}
\end{equation}
To extend the metric (\ref {43.17}) outside of region (\ref {43.21})
one uses the Penrose-Khan trick and substitutes in (\ref {43.17})
\begin {equation} 
                                                          \label {43.22}
u\to u\theta (u),~~v\to v\theta (v),
\end   {equation} 
\begin{equation}
\theta (x)=\left\{
\begin{array}{cc}
	1, ~& x>0  \\
	0, ~& x<0
\end{array}
\right.
	\label{6.3}
\end{equation}
Fig.1 illustrates this metric.
The region {\bf I} ($u<0,~~v<0$) is Minkowskian.
Regions {\bf II} and {\bf III} contain the approaching plane
waves with the following metrics
\begin {equation} 
                                                          \label {43.23}
(ds^{II})^{2}= 4X(u)dudv-\Omega (u)[\chi (u) dy^{2}+\frac{1}{\chi (u)}
(dx-\lambda (u)dy)^{2}],
\end   {equation} 
\begin {equation} 
                                                          \label {43.24}
(ds^{III})^{2}= 4X(v)dudv-\Omega (v)[\chi (v) dy^{2}+\frac{1}{\chi (v)}
(dx-\lambda (v)dy)^{2}],
\end   {equation} 
where
\begin {equation} 
                                                          \label {43.25}
X(u)=(1-2p\sin u)+\sin ^{2}u,~~\Omega (u) =\cos^{2}u,
\end   {equation} 
\begin {equation} 
                                                          \label {43.26}
\lambda (u)=2q(\sin u -\frac{1}{1+p})
\end   {equation} 
In (\ref {43.23})
\begin{equation}
u<\pi/2,~~v<0,
                                                          \label{43.27}
\end{equation}
and in (\ref {43.24})
\begin{equation}
u<0,~~v<\pi/2,
                                                          \label{43.28}
\end{equation}
The region {\bf IV} is the interaction region with metric (\ref {43.17}).

\section{Semiclassical Transition Amplitude}
\setcounter{equation}{0}

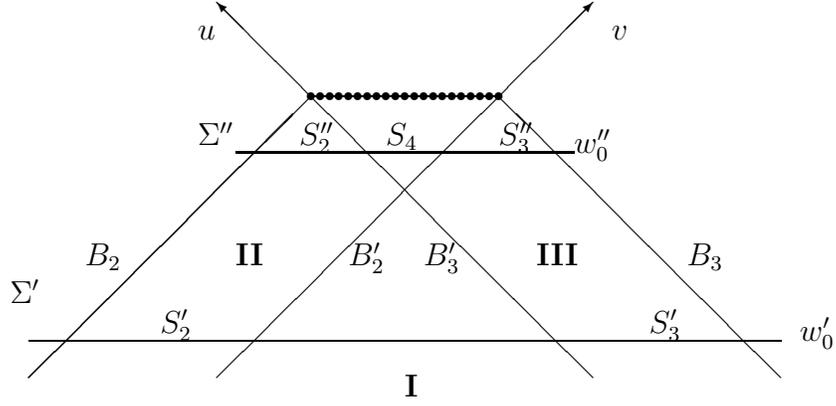
\begin{figure}
\setlength{\unitlength}{0.5cm}
\begin{center}
\begin{picture}(8,10)(-5,-5)
  \put(-2.5,2.5){\line(1,0){5.0}}
  \multiput(-2.5,2.5)(0.25,0.0){21}{\circle*{0.2}}
\put(-5.0,-5.0){\vector(1,1){10.0} }
\put(5.0,-5.0){\vector(-1,1){10.0} }
        \put(-10.0,-5.0){\line(1,1){7.5}}
 \put(-10.0,-5.0){\line(1,1){7.0}}
\put(-10.0,-4.0){\line(1,0){20.0}}
\put(-4.5,1.0){\line(1,0){9.0}}
\put(10.0,-5.0){\line(-1,1){7.5}}
 \put(5.5,4.0){\mbox {$v$}}
 \put(-5.5,4.0){\mbox {$u$}}
 \put(-5.5,1.2){\mbox {$\Sigma ''$}}
 \put(4.5,1.0){\mbox {$w_{0} ''$}}
 \put(-0.5,1.2){\mbox {$S_{4}$}}
 \put(2.5,1.2){\mbox {$S''_{3}$}}
 \put(-2.8,1.2){\mbox {$S''_{2}$}}
\put(-10.5,-3.0){\mbox {$\Sigma '$}}
 \put(10.5,-4.0){\mbox {$w_{0}'$}}
\put(6.5,-3.8){\mbox {$S'_{3}$}}
  \put(-6.5,-3.8){\mbox {$S'_{2}$}}
  \put(-8.5,-2.0){\mbox {$B_{2}$}}
 \put(-1.5,-2.0){\mbox {$B'_{2}$}}
  \put(7.5,-2.0){\mbox {$B_{3}$}}
 \put(0.5,-2.0){\mbox {$B'_{3}$}}

\put(-4.5,-2.0){\mbox {{\bf II}}}
\put(3.5,-2.0){\mbox {{\bf III}}}
\put(0.0,-5.5){\mbox {{\bf I}}}
\end{picture}
 \end{center}
\caption{Initial and final Cauchy surfaces}\label{f5}
\end{figure}

In this section we  study the transition amplitude
\begin {equation} 
                                                          \label {5.1}
<2pw, WH|2pw>
\end   {equation} 
from a state $|2pw>$ of two plane gravitational waves to
a state $|2pw,WH>$ containing white-hole and two plane gravitational waves.
Analogous calculations may be performed for the transition  amplitude
\begin {equation} 
                                                          \label {5.2}
<2pw|2pw>
\end   {equation} 
from two plane gravitational waves back  to two plane gravitational waves
and for the amplitude
\begin {equation} 
                                                          \label {5.3}
<BH|2pw>
\end   {equation} 
for a process 2 plane waves $\to$ black holes.
We can consider these two amplitudes as amplitudes of two independent
channels.

To find these transition amplitudes in the semiclassical approximation
according to (\ref{3.12}),
we have to evaluate the boundary term for the classical solution interpolating
between two plane gravitational initial state and
an appropriate final state.

Let us start considering  the transition (\ref {5.1}).
The corresponding classical solution is shown in Fig.2.
In this case the initial Cauchy surface $\Sigma '$ crosses regions
{\bf I},{\bf II} and {\bf III} and the final surface crosses regions
{\bf II},{\bf IV} and {\bf III}.   For the  surfaces
$\Sigma '$ and $\Sigma ''$
shown in Fig.2 equation (\ref {3.10}) reads
\begin{equation}
\Sigma ':  \sigma = u+v-w'_{0} =0, ~w'_{0}\leq 0
                                                         \label{7.1}
\end{equation}
 \begin{equation}
\Sigma '': \sigma = u+v-w''_{0}=0, ~0<w''_{0}\leq \pi /2
                                                         \label{7.2}
\end{equation}
In this case the other parts of the boundary defining the boundary term
are given by equations
  \begin{equation}
 B_{2}: ~u-\pi/2 =0,  ~w'_{0}-\pi /2 \leq v\leq  w''_{0}-\pi /2;~~
 B_{2}': ~u=0,~ ~w'_{0}-\pi /2 \leq v\leq  0
	\label{7.3}
\end{equation}
  \begin{equation}
 B_{3}: ~v-\pi/2 =0,  ~w'_{0}-\pi /2 \leq u\leq  w''_{0}-\pi /2;~~
 B_{3}': ~v=0,~ ~w'_{0}-\pi /2 \leq u\leq  0
	\label{7.4}
\end{equation}

To calculate the phase factor let us give expressions for $f^{\mu}$
on the boundary.
In Minkowski space we have  $f^{\mu}=0$. In all regions
$f^{x}=f^{y}=0$, moreover
 \begin{equation}
f^{u}|_{II}=0,  ~~~f^{v}|_{III}=0
                                                   \label{7.5}
\end{equation}
So the value of the action (\ref{3.11})
for the two plane waves solution (\ref{43.23}), (\ref{43.24})
is reduced to the sum of two terms each of which represents a contribution
from the Cauchy surface $\Sigma '$ and $\Sigma ''$, respectively,
\begin{equation}
S(g_{cl}^{(2pw)})=S^{\Sigma '}+S^{\Sigma ''},
                                                     \label{7.6}
\end{equation}
where
\begin{equation}
S^{\Sigma '}= S'_{2}+S'_{3},~~S^{\Sigma ''}= S''_{2}+S''_{3}+ S''_{4},
	\label{7.7}
\end{equation}
see Fig.2. So
\begin {equation} 
                                                          \label {7.71}
<2pw|2pw, WH>={\cal Z}\exp \{\frac{i}{\hbar }(S'_{2}+S'_{3}
-S''_{2}-S''_{3}- S''_{4})\}
\end   {equation} 
here ${\cal Z}$ is the one-loop contribution.

Taking into account the special
form of the metric one can write $f^{w}$
in  regions {\bf II}, {\bf III} and {\bf IV} respectively in the
following form
\begin{equation}
\sqrt {-g}f^{w}|_{II}=\Omega (u)\partial _{u}\ln
[X(u)\Omega ^{2}(u)]
                                            \label{7.8}
\end{equation}
 \begin{equation}
\sqrt {-g}f^{w}|_{III}=\Omega (v)\partial _{v}\ln
[X(v)\Omega ^{2}(v)]                           \label{7.9}
\end{equation}
\begin{equation}
\sqrt {-g}f^{w}|_{IV}=\Omega (w,z)\partial _{w}\ln
[X(w,z)\Omega ^{2}(w,z)]                           \label{7.10}
\end{equation}
$$w=u+v, ~~z=v-u.$$
The representation (\ref {7.8}) was obtained in the following way.
Metric (\ref {43.17})
in $(w,z,x,y)$  coordinates admits the representation
\begin{equation}
g_{\mu\nu}=\left (
\begin{array}{cc}
        g_{\alpha \beta} ~& 0  \\
        0 ~& g_{ij}
\end{array}
\right ).
        \label{7.101}
\end{equation}
with the diagonal matrix $g_{\alpha \beta}$,
\begin{equation}
g_{\alpha \beta}=\left (
\begin{array}{cc}
        g_{ww} ~& 0  \\
        0 ~&g_{zz}
\end{array}
\right ). ,   ~~g_{ww}=-g_{zz}=X
        \label{7.102}
\end{equation}
and  $g_{ij}$ of the form
\begin{equation}
g_{ij}=\Omega \tilde {g_{ij}}                           \label{7.103}
\end{equation}
where
\begin{equation}
                                                        \label{7.104}
  \tilde {g_{ij}}
= \left (
\begin{array}{cc}
     \frac{1}{\chi}  & -\frac{\lambda}{\chi}  \\
        \frac{\lambda}{\chi}& \chi +\frac{\lambda ^{2}}{\chi}
\end{array}
\right ).
\end{equation}
being an element of the group $SL(2)$.
{}From
equation  (\ref {3.3}) it follows
\begin {equation} 
                                                          \label {7.105}
f^{w}=g^{ww}(g^{zz}\partial _{w} g_{ww}+g^{ij}\partial _{w} g_{ij})
\end   {equation} 
Taking into account that the trace  $tr\tilde
{g}^{-1}\partial\tilde {g}=0$ for the group $SL(2)$, we get
\begin {equation} 
                                                          \label {7.106}
f^{w}=g^{ww}(g^{zz}\partial _{w} g_{ww}+2\Omega ^{-1}\partial _{w} \Omega ),
\end   {equation} 
from which follows the representation (\ref {7.10}).

We have to compute the value of the boundary term in the action on the plane
wave solution. Because the plane wave metric does not depend on $x$ and
$y$ coordinates we have to introduce a cut-off $L$ in these directions.
Simple calculations give
\begin{equation}
S_{2}'=-\frac{1}{16\pi G}\int^{L/2}_{-L/2} dx\int^{L/2}_{-L/2} dy
\int ^{0}_{\pi/2}du\int ^{w'_{0}}_{w'_{0}-\pi /2}
dv \delta (u+v -w_{0}')\Omega (u)\partial _{u}\ln
[X(u)\Omega ^{2}(u)]
	\label{7.11}
\end{equation}
$$=-\frac{1}{16\pi G}L^{2}\int ^{0}_{\pi/2}du\cos ^{2}u\partial _{u}[\ln
((1-p\sin u )^{2}+
q^{2}\sin ^{2}u)\cos ^{4}u]
$$
$$=-\frac{2L^{2}}{16\pi G}(\int ^{0}_{1}du
\frac{(1-u^{2})(u-p)}{1-2up +u^{2}}-4\int ^{0}_{1}udu)$$
$$=-\frac{L^{2}}{16\pi G}(-3-2p+2q^{2}\ln 2(1-p)+4pq \arctan \frac{q}{1-p}) $$
and
\begin{equation}
S_{2}''=-S'_{2}+\bar {S}''_{2},        \label{7.12}
\end{equation}
where
\begin {equation} 
                                                          \label {7.121}
\bar {S}''_{2}=
\frac{L^{2}}{16\pi G}
\int ^{0}_{w''_{0}}du\cos ^{2}u\partial _{u}[\ln ((1-p\sin u )^{2}+
q^{2}\sin ^{2}u)\cos ^{4}u]
\end   {equation} 
$$=\frac{L^{2}}{16\pi G}[-3\sin ^{2}w''_{0}-2p\sin w''_{0}+2q^{2}\ln (
\sin ^{2}w''_{0}-2p\sin w''_{0}+1)+4pq \arctan \frac{q\sin w''_{0}}
{1-p\sin w''_{0}}] $$

One can take the box $L^{2}$ in the $x-y$ plane and consider the periodic
boundary conditions. If $g_{\mu \nu}$ is a solution
of the vacuum Einstein equations then $\tilde {g}_{\mu \nu}$$=
\frac{c}{L^{2}}g_{\mu \nu}$ is also a solution. $c$ is some constant
of dimension of square of the length, in particular, we can take $c=m^{2}$.
If we take metric $\tilde {g}_{\mu \nu}$ in our computations of the transition
amplitude then $L^{2}$ in (\ref {7.11}) and (\ref {7.121}) disappears.

The part of the boundary which cross the white hole region gives the
following contribution
\begin{equation}
S''_{4}=-\frac{1}{16\pi G}\int^{L/2}_{-L/2} dx\int^{L/2}_{-L/2} dy
\int ^{w'_{0}}_{-w'_{0}}dz\Omega (w,z)\partial _{w}\ln
[X(w,z)\Omega ^{2}(w,z)]
	\label{7.13}
\end{equation}
$$=-\frac{L^{2}}{16\pi G}\int ^{-w''_{0}}_{w''_{0}}dz
[\frac{-2p(1-p\sin w''_{0})\cos ^{2}w''_{0}
\cos z}{(1-p\sin w''_{0})^{2}+q^{2}\sin ^{2}z
}-2\sin  w''_{0}\cos z ] $$
Now let us consider the special case when $ w''_{0}=\pi/2$.
In this case we do not have plane waves in the final state
but only a white hole. Then  only the second term in the RHS of
(\ref {7.13}) survives and  we get
$$S^{\Sigma ''}= S''_{4}=\frac{L^{2}}{4\pi G}.$$
Therefore the amplitude describing the transition from two plane waves to
white hole in the semiclassical approximation is given by
\begin {equation} 
                                                          \label {7.15}
<WH|2pw>={\cal Z}\exp \{\frac{i}{ \hbar }(S'_{2}+S'_{3}
- S''_{4})\}=
\end   {equation} 
$${\cal Z}\exp \{-\frac{iL^{2}}{8\pi G \hbar }[
-1-2p+2q^{2}\ln 2(1-p)+4pq \arctan \frac{q}{1-p}]\}$$

To get the semiclassical answer for the transition amplitude
$<2pw|2pw>$
we have to consider the classical solution
which describes the metric extended  as shown in Fig.4.
We see from equations (\ref {7.11}) that $S'_{2}$ does not depend on
$w_{0}'$ and therefore for the transition amplitude describing the elastic
scattering of two plane waves the phase factors  is zero, and
\begin {equation} 
                                                          \label {7.16}
<2pw|2pw>={\cal Z}
\end   {equation} 

In this section we have computed only the phases of transition amplitudes.
However they contain a nontrivial information about our processes because we
different channels available and there would be a quantum mechanical
interference
between them.

$$~$$
{\bf CONCLUSION}
$$~$$
We have considered a possible mechanism for black hole creation in the
 collision  of two light particles
at planckian energies. Many questions deserve further study.
In particular there is an important question of how to relate the momenta
of colliding particles with characteristics of colliding plane waves.
It is known that a plane wave does not lead to polarization of vacuum
\cite {Gibb}. It was shown that a plane wave is an exact string background
\cite {AK,NWi,THo,Tse} and there is a duality with extreme black hole
\cite {Rena}. One can conjecture that a dilaton gravity analogue
of the CFX duality between colliding plane waves and non-extremal black holes
discussed in this paper could lead to a corresponding string duality.
$$~$$
{\bf ACKNOWLEDGMENT}
$$~$$
This work has been supported in part by
an operating grant from the Natural Sciences
and Engineering Research Council of Canada.
I.A. and I.V. thank the Department of Physics for kind hospitality
during their stay at Simon Fraser University.
I.A. and I.V. are supported in part also by
International Science Foundation under the grant M1L300.
I.V. is grateful to  D.Amati, A.Barvinsky, W.Israel,
V.Frolov, D.Page, G.Veneziano and  A.Zelnikov for stimulating
and critical discussions.
$$~$$

\end{document}